\documentstyle[]{article}
\newcommand{\be}{\begin{equation}}
\newcommand{\ee}{\end{equation}}
\begin{document}
\title{The Quark-Hadron Phase Transition,
QCD Lattice Calculations and
Inhomogeneous Big-Bang Nucleosynthesis}
\author{{\bf A.A. Coley and T. Trappenberg} \\
\vspace*{.5cm}
Department of Mathematics, Statistics
and Computing Science \\
Dalhousie University \\
Halifax, N.S., B3H 3J5, Canada}
\maketitle
\begin{abstract}
We review recent lattice QCD results for the surface tension at the
finite temperature quark-hadron phase transition and discuss their
implications on the possible scale of inhomogeneities.
In the quenched approximation the average distance between
nucleating centers is smaller than the diffusion length of a protron,
so that inhomogeneities are washed out by the time nucleosynthesis sets in.
Consequently the baryon density fluctuations formed by a QCD phase transition
in the early universe cannot significantly affect standard big-bang
nucleosynthesis calculations and certainly cannot allow baryons to close the
universe.
At present lattice results are inconclusive when
dynamical fermions are included.
\end{abstract}
\pagebreak
\section{Introduction}
The density of luminous matter is $\Omega_L \sim 0.01$
(to within a factor of two) and the density
of clustered matter needed to account for the stability of galaxies or
large-scale motions is  $\Omega_G \sim 0.1 - 0.2$.
Also, inflation predicts that $\Omega \sim \Omega_c = 1$ ,
where $\Omega_c$  is the critical density for closure  ($\Omega$ is the
ratio of the present density to the critical density of the universe).
However, standard (homogeneous density) big-bang nucleosynthesis (BBN)
calculations constrain the ratio of the present baryon density to
$\Omega_c$  by  $0.01 \leq \Omega_B \leq 0.15$ \cite{A}
(where uncertainties in the value of the Hubble parameter are included).
While the value of $\Omega_B$ is consistent with $\Omega_G$, if
$\Omega \sim 1$ then the majority of matter in the universe is non-baryonic
(consisting of, for example, axions, WIMPS, massive neutrinos, etc.).

Recently the physics of the quark-hadron phase transition has been of
much interest (cf. Reeves \cite{B} and references therein).
Of particular interest, from an astrophysical point of view, is
the suggestion that (large) baryon density fluctuations are formed
by QCD phase transitions which occur in the early universe \cite{C},
that will then affect BBN calculations \cite{D,D2}.  In particular,
initial calculations indicated that the upper bound on
$\Omega_B$ from BBN might be relaxed and it was
suggested that $\Omega_B \sim 1$ might be reconciled with observations
\cite{Ea,Eb}.  We shall discuss inhomogeneous BBN in section 2.

However, many parameters of the quark-hadron phase transition are poorly
known, and extensive QCD calculations on networks are necessary to
determine the affects of inhomogeneous BBN more accurately.
In section 3 recent results from QCD lattice calculations will be reviewed.
The consequences for BBN and the constraints on $\Omega_B$ will be briefly
discussed in the concluding section.

\section{Primordial Nucleosynthesis}
The cosmological implications of the quark-hadron phase transition arises
from the fact that
it may induce baryon density inhomogeneities in the cosmic fluid,
thereby producing
inhomogeneities in the ratio of neutrons to protons which then modify
the abundances of the light
elements obtained in the standard (homogeneous) BBN calculations.
Larger values of $\Omega_B$  than
in the baryon-homogeneous cosmologies are allowed,
since in both proton-rich and neutron-rich
regions the burning of neutrons and protons through D up to $^4$He is
less efficient (for larger baryon densities) and at later times
(when sufficient neutrons have decayed) the temperature is
lower and the subsequent burning of  D  is less efficient
(than in the homogeneous case in both cases) resulting in relatively less
$^4$He  being made in both neutron and proton-rich regions but
relatively greater amounts of D (and other heavier elements) being made
in the neutron-rich regions.

Assuming that the phase transition is first order,
the new phase is not reached immediately; the system is cooled through
the critical temperature  $T_c$ , but overcooling occurs in
which fluctuations create small volumes of the new phase \cite{KaKu}.
The nucleated bubbles of the hadronic phase then expand in the quark-gluon
phase with the velocity of light in the quark-gluon plasma $v_s=1/\sqrt{3}$.
The shock wave will then reheat the plasma, so that no further
nucleation occurs.
The scale of the inhomogeneities produced at the quark-hadron phase transition
within this simple bubble nucleation scenario
is given by the mean separation of the nucleating centers $l$.
Assuming that the phase transition occurs within thermodynamic equilibrium,
the mean distance between nucleating centers has been estimated by Meyer
{\it et al.} \cite{H} to be (in terms of the inverse Hubble time $H$)
\be
l \times H \approx 4.38 \frac{(0.4)^3 \sqrt{3}}{32 \pi}
[\frac{\alpha}{ T_c^3}]^{3/2}  [\frac{L}{ T_c^4}]^{-1} ,
\ee
which depends on the QCD parameters
$L$ (the latent heat) and $\alpha$ (the surface tension, which is
the surface energy density of the hadronized bubbles).
With the inclusion of the quark degree of freedoms the
temperature dependent Hubble time was estimated to be \cite{BaGa92}
\be
H^{-1}=\frac{3}{\sqrt{164 \pi^3}} \frac{M_{Pl}}{T^2} ,
\ee
with the Planck mass $M_{Pl}=2 \times 10^{19} GeV$,
so that the scale of inhomogeneities can be further reduced to
\be
    l \approx 8 \times 10^5 m [\frac{\alpha}{ T_c^3}]^{3/2}
[\frac{L}{ T_c^4}]^{-1} [T_c/(MeV)]^{-2} .
\ee
QCD estimates of  $T_c$  indicate that  $T_c = 200 \pm 50 MeV$
in the quenched approximation of QCD \cite{G} and $T_c = 80 \pm 20 MeV$
with four flavours of light quarks \cite{MTc}.
Estimates for $l$ have been discussed by various authors \cite{H,I,F}, although
within  the uncertainties of the QCD parameters in the original calculations
$l$ could range, in principle, from zero to relatively large values.
A better estimate for $l$ must therefore come from detailed
quantum lattice theory calculations.

The universe cools from the quark-hadron phase transition at  $T_c$
until the time of BBN at $T_N$  ($t_N \sim 1$sec),
by which time antimatter has disappeared.
% and the important
%parameter is the contrast $R$ in the baryonic number density
%(between the hadronized bubbles
%and the quark sea from which they were formed).
%Approximations of $R$ indicate that it is
%essentially a function of  $T_c$ ;
%calculations of $R$ (using different models) \cite{D,D2,Ea,J} are in
%agreement for  $T_c < 150 MeV$, with $R \sim 100$ for  $T_c \sim 100 MeV$,
%but disagree for higher  $T_c$  with  $1 < R < 10$.
%It has been argued \cite{H} that the best estimates for  $T_c$,
%$180 < T_c < 220 MeV$, lead to values of  $5 < R < 10$
%at chemical potential equilibrium, although the effective value of $R$
%should be somewhat higher (but not much).
%Potential constraints on the relative sizes of the
%parameters $l$ and $R$  have also been discussed \cite{I}.
Before $t_N$ the neutron-to-proton ratio is governed by weak interactions
(and given by the Boltzmann formula), but for $t > 1$sec
the weak processes are no longer in thermal equilibrium.
Neutrons diffuse from high baryon-density phases to low-density phases,
changing both their density and their neutron-to-proton ratio, so that
by $T_N$ high-density proton-rich regions and
lower-density neutron-rich regions would exist, thereby affecting
nucleosynthetic yields \cite{D}.
The extent of neutron diffusion is a function of $l$ (and the
fractional volume of the high-density phase, $f_v$).
The smaller the value of  $l$ the less the effect of inhomogeneities,
and so for significant changes to nucleosynthetic yields a large value of
$l$ is necessary.
For example, within a typical model one would
need an $l$ in the range of $l \sim 5-150 m$ to get $\Omega \geq 0.15$,
( cf. \cite{H}).

Many authors \cite{D,D2,Eb,H,I,K,L} have estimated numerically the
nucleosynthetic yields based on various physical models and approximations
(e.g., 2-64 zone models, whether or not neutron and proton diffusion is
taken into account before BBN, and so on) and
various values for the QCD parameters, particularly to determine whether the
observed abundances of the light nuclides  D, $^3$He, $^4$He,
$^7$Li \cite{A,M} can be reconciled with a critical baryon density
$\Omega_B \sim 1$.
Although the various studies are in good agreement with each other,
the ignorance of the exact values of many QCD
parameters lead to corresponding uncertainties in the results.

%Essentially, if $l< 0.1$ 1h  the proton diffusion becomes important and
%computations will yield standard (homogeneous-density) nucleosynthetic
%yields, whereas for large  $l$ ($l \sim 10^4 $1h) neutron diffusion could
%not take place before $t_N$ giving rise to the same results as in the
%inhomogeneous-density standard model.
%Clearly  $l \lsim 10 cm$  for important cosmological effects to result.

%For the nucleosynthesis the important
%parameter is the contrast $R$ in the baryonic number density
%(between the hadronized bubbles
%and the quark sea from which they were formed).
%%Approximations of $R$ indicate that it is essentially a function of  $T_c$ ;
%In Reeves \cite{B} it was pointed out that higher baryonic densities can
%only be achieved with larger values of $R$  (e.g.,
%$\rho_B \gsim 10\times10^{-31}gm \;\; cm^{-3}$ for $R \gsim 10$,
%$\rho_B \gsim 20\times10^{-31}gm \;\; cm^{-3}$ for
%$R > 100$), and for the best estimates of the QCD parameters available
%at that time he concluded a reasonable allowable range for
%$\rho_B$ to be $2-10\times10^{-31}gm \;\; cm^{-3}$,
%which is not large enough to allow baryons to close the universe.
%Moreover, Reeves argued \cite{B} that if it can be shown that if
%$l < 1  $ 1h or if  $R \lsim 5$  (or if $f_v < 0.1$) ,
%then non-baryonic matter will be required for $\Omega \sim 1$.
%Clearly more improved values of the QCD parameters $l$ and $R$
%(and  $T_c$  and $f_v$) are needed from lattice calculations.

\section{Calculations from lattice QCD}

Due to the lack of appropriate models and/or experimental knowledge of the
quark-hadron transition, thermodynamic parameters of the transition have to
be calculated directly from the underlying theory, which means that one has
to consider QCD at the transition temperature.
At very high physical temperature QCD can be treated by perturbation theory
because the separation of quarks will be small and the effective coupling
between quarks and gluons will be governed by the asymptotic freedom of QCD.
However, as has been pointed out by Linde \cite{Li80}, perturbation theory will
brake down at $O(g^6)$ (here $g$ is the QCD coupling constant) due to
the infrared problem of the massless Yang-Mills theory.
Therefore nonperturbative calculational techniques such as Monte Carlo
simulations of the latticized theory are the only known way to get reliable
information on the hadron transition.

After discretizing the QCD action on a 4-dimensional hypercubic lattice the
theory can be quantized due to Feynman's path integral formalism,
where the lattice cutoff $\Lambda = \frac{1}{a}$ ($a$ is the lattice
spacing) regularizes the infrared singularities \cite{Wilson,Rothe}.
The path integral is then formally equivalent to a partition function
in statistical mechanics and can be treated by nonperturbative methods.
To remove the cutoff the continuum limit has to be performed at a second
order phase transition of the equivalent statistical mechanical model,
where the relevant scale on the lattice, the correlation length, diverges.
In practice one can not reach this limit in simulations, but this is also
not necessary because an ansatz of asymptotic scaling is enough to
extrapolate reliably with the renormalization group to the continuum limit.

The most general method to calculate expectation values of the latticized
theory is the numerical Monte Carlo simulation method. There, in addition
to the discretization, the space-time dimensions have to be restricted
to a finite volume. The difficulties in lattice simulations are therefore to
keep the lattice volumes large enough compared to the correlation length, so
that finite size effects are under control with finite size scaling theories.

Finite physical temperatures can be taken into account (as in perturbation
theory) by taking finite lattice extensions in the time-direction.
In finite temperature lattice calculations, however, the number of lattice
points $N_t$ in the time direction have to be large enough to
be in the asymptotic scaling region of the observables under investigation.
Whether this can be realized is mainly a question of the available computer
time and how fast the scaling of the observables sets in.

Monte Carlo simulations of lattice QCD have established a phase transition
between a chiral symmetric quark-gluon plasma phase at high physical
temperatures with no color confinement and the hadron phase at low
temperatures where quarks are confined in hadrons and the chiral symmetry
is broken spontaneously as proposed by McLarren and Svetitzky
\cite{McSv81} (see for example \cite{Rothe} for details).
The lattice calculations have established a first order phase transition in the
quenched approximation, where the quarks are infinitely heavy and therefore
only the gauge degrees of freedom have to be treated dynamically.
Also, in the case of at least four flavours of dynamical light fermions
lattice results indicate a weak first order phase transition \cite{MTc}.
Therefore, as discussed above, there exists the possibility that hadronization
takes place through nucleation of hadronic droplets in a supercooled plasma.
In addition, since nucleation is a dynamical process it is governed by
equilibrium parameters of the transition such as the surface free energy
and the latent heat, which can be calculated in lattice QCD.

\subsection{The interface tension in quenched QCD}

The interface tension is given by the free energy of the interface between
coexisting phases normalized to the area of the interface \cite{LaLi},
\be
\alpha = \frac{F}{A} ,
\ee
and is usually given in units of the critical temperature $T_c$.
The free energy is proportional to the logarithm of the partition
function and the free energy of the interface accounts only for the
small difference of the free energies of bulk phases.
Therefore, the interface tension is difficult to calculate in Monte Carlo
simulations. Reliable results have
been derived until now only in the quenched approximation, on which we will
concentrate in this subsection. The influence of dynamical fermions will be
discussed in the next subsection.

Several methods have been used over the past three years to extract the
surface tension in quenched QCD. Most of the results are obtained on lattices
with the small time extent $N_t=2$, but the first results on
$N_t=4$ and $N_t=6$ lattices are now being quoted.
We have collected recent results for the surface tension together in Table~1.
% here comes table 1
\begin{table}
\begin{center}
\begin{tabular}{llll}
        &          & $\alpha / T_c^3$ \\ \hline
        &          &          &             \\
ref.    &  $N_t=2$   &  $N_t=4$   & $N_t=6$       \\ \hline
23      &   0.08(2)  &            &             \\
24,25   &   0.12(2)  &  0.027(4)  &             \\
26      &   0.139(4) &            &             \\
28      &   0.092(4) &  0.025(4)  &             \\
30      &            &  0.029(2)  & 0.022(4) \\
\hline
\end{tabular}
\end{center}
\caption{\em{The surface tension in quenched QCD
             calculated with different methods
             on lattices with different time extent $N_t$
.}}
\end{table}
Kajantie {\it et al.} \cite{Helsinki} have
used a surface tension operator derived from the partition function
by derivation of the partition function with respect to the area of the
interface. In this method some parameters have to be estimated by perturbative
methods and thus reliability can be questioned at the transition temperature.

Potvin {\it et al.} \cite{Boston1,Boston2} have
used integral methods, where the average action is integrated in the space
of the gauge coupling parameter. The interface is thereby generated by a
temperature difference of two parts of the lattice or by applying different
external fields to the two sublattices. The main difficulty in this principally
exact method is the extrapolation of the numerical results to the zero
difference limit of the temperatures
or the external fields of the sublattices.
This method has been recently used to calculate the surface tension in
simulations of $N_t=4$ lattices \cite{Boston2}.

An elegant transfermatrix method has been applied to quenched QCD in
\cite{GrLa93}. In this method finite size effects of the spectrum of states
are used to extract the surface tension. With this method very precise
results have been obtained for $N_t=2$ lattices; however, due to the lack of
global updating algorithms it has not been applied to lattices with larger
time extents.

The most recent results have been derived by an histogram method
proposed some time ago by Binder \cite{Binder}, where finite size effects in
histograms of different operators due to interface effects
are used to calculate the surface tension. The reason
for a renaissance of this method is that the recently proposed multicanonical
algorithm produces more statistics of interface configurations and that the
analysis method is easy to apply to existing high statistical data of the
QCDPAX collaboration. Grossmann and Laursen \cite{Binderm}
have analysed multicanonical data on $N_t=2$ lattices with
finite size formulas, where fluctuations of the interface have been
taken into account by a capillary wave model and interfacial interactions
have been reduced by the use of rectangular lattices. These authors have
also analysed $N_t=4$ data of the QCDPAX collaboration \cite{QCDPAX}
with the capillary wave improved finite size formulas.
Preliminary analysis of the QCDPAX data by Iwasaki {\it et al.} \cite{leo}
lead to preliminary estimates for $N_t=6$ data also.

Although the scattering of the $N_t=2$ results show how difficult the
extraction of the surface tension is, the data indicate at least the
order of magnitude of $\alpha$
to be expected at the finite temperature phase transition of gluonic
matter. The decrease of the values with increasing $N_t$ indicate, however,
that the region of asymptotic scaling has not yet been reached.

Lattice results for the latent heat at the finite temperature phase transition
in quenched QCD can be estimated from \cite{QCDPAX} to be
$L/T_c^4 = 3.10(6)$ on the $N_t=4$ lattice and $L/T_c^4 = 1.98(5)$ on the
$N_t=6$ lattice.
With the largest value of the surface tension on $N_t=2$ lattices
together with the lowest estimate of $150 MeV$ for the transition temperature
and $L/T_c^4=1.98$ for the latent heat, we obtain
an estimate for an upper bound on the average distance between nucleating
centers (using formula (3)) of $l \approx 0.93 m$. However, with
the estimate of $\alpha/T_c^3=0.025$ on $N_t =4$ lattices, which is
closer to the continuum limit, and with the same values of $L$ and $T_c$
as used above, we obtain the more realistic estimate of $l \approx 0.07 m$.
Therefore the average distance of nucleating centers is likely to be smaller
than the diffusion length of protons ($\sim 0.5 m$) and
neutrons ($\sim 30 m$), so that inhomogeneities created
at the finite temperature phase transition of pure gluonic matter
are washed out by the time primodial nucleosynthesis sets in.
These conclusions are consistent with those of Ref.~\cite{BaGa92}.

Also a large fraction of supercooling seem to be unlikely in quenched QCD.
At the finite temperature phase transition of the pure SU(3) gauge theory
three different deconfinement phases can coexist
with the confinement phase, because
the Z(3) center symmetry of the SU(3) gauge symmetry is
spontaneously broken at $T > T_c$  \cite{McSv81}.
As pointed out in \cite{GrLa93} it is then most likely that a layer of
the confined phase will completely wet the network of interfaces between
deconfinement phases, which is rather different to the bubble nucleation
scenario. However, this is only true in pure gluonic matter.

\subsection{The influence of dynamical fermions on the hadronization}

Although the quenched approximation of QCD can give a rough idea of the
behavior of strongly interacting matter, the influence of the fermion
dynamics has to be taken into account in a real world QCD calculation.
Unfortunately the simulation of dynamical fermions is
not only beset with technical difficulties but is also very computer
time consuming, so that these studies
have until now not reached the quality of quenched simulations.

Although it has not yet been demonstrated that there exists a finite
temperature phase transition with a realistic fermion spectrum
(see \cite{BrBu90} for a discussion), it is now widely accepted that in
full QCD a weak first order phase transition also exists with at least
two light and one not-too heavy quark flavours.
The transition temperature is, however, flavour dependent and drops
from the above quoted $200 MeV$ in the pure gauge theory
to $T_c \approx 100 MeV$ for four degenerate quark flavours \cite{MTc}.
The quark degrees of freedom also breaks the Z(3) symmetry of the pure
SU(3) Lagrangian, so that complete wetting will no longer inhibit
supercooling. It has been estimated by Ignatius {\it et al.} \cite{Ig}
that a metastable Z(3) phase will convert into a stable one by
a temperature of $T\approx 10 TeV$ around $10^{-13}$ sec after the
big bang.

Lattice results for the surface tension with dynamical fermions are
not yet available. The only simulations of Markum {\it et al.}
\cite{vienna} lead to an
inconsistent negative value of the surface tension and show only
that it seems not to be considerably larger than in the quenched approximation.

On the other hand, the transition is weaker than in the quenched approximation
leading to a larger correlation length and a smaller latent heat. Together
with the reduction of the critical temperature, these factors might lead to
a larger average separation of nucleating centers
as indicated by Eq.(3). Whether these would result in
significant inhomogeneities by the time primordial nucleosynthesis sets in
is unclear however.
More calculations on the lattice with dynamical fermions need to be done.

\section{Conclusions}
Original studies, based on a simple two-zone model in which back
diffusion of neutrons are neglected during nucleosynthesis suggested
that a $\Omega \sim 1$  model might be consistent with
light element observations.
The importance of neutron diffusion during nucleosynthesis was
emphasised \cite{Eb}, leading to numerical codes treating nucleosynthesis
and diffusion simultaneously \cite{I,F,N}.
Neutron diffusion also leads to an overestimation of the neutron number
density in the low density-region in a simple two-zone model, resulting in
more realistic multi-zone calculations \cite{I,F}.
These calculations led to upper limits on $\Omega_B$ lower than the
critical value.
Finally, a large nuclear reaction network including very neutron rich nuclei
and heavy elements were included in the multi-zone calculations \cite{N}
(using various values of the QCD parameters), leading
to a considerably lower upper limit on $\Omega_B$ and ensuring that an
$\Omega_B \sim 1$  model is inconsistent with the observed abundances of
the light elements even if the universe is inhomogeneous.
Recently \cite{O} the yields of primordial light elements from
baryon-inhomogeneous BBN have been recalibrated using new 'improved'
diffusion coefficients calculated from relativistic kinetic theory,
with the result that the yields previously obtained using more crudely
derived diffusion coefficients (e.g., Refs. \cite{D2,Ea,Eb,H,I,L,F})
remain virtually unchanged.
The various results are essentially consistent and in basic agreement
with Reeves' \cite{B} quoted limits of $0.01 < \Omega_B < 0.2$
(a range which is somewhat larger than that in the standard BBN scenario)
based on the then allowable ranges of the QCD parameters.

To date there are no lattice calculations available with a realistic
fermion spectrum close to the continuum limit, so that
definitive conclusions for the quark-hadron transition cannot be
made yet. Further research in this field is necessary to get a better
understanding of the phase transitions in the early universe.
However, {\it in the models studied so far there is no evidence at all
for a strong supercooling and a cosmologically
relevant scale for the inhomogeneities created at the quark-hadron transition}.

The most detailed studies of the surface tension have been done in
the quenched approximation of QCD, where the fermion dynamics have been
neglected. In this approximation only a small surface tension was found.
Although the results are still not in the asymptotic scaling region close to
the continuum limit, the trend is that for larger lattices even smaller
values of the surface tension result, leading to scales of inhomogeneities
which are irrelevant for primordial nucleosynthesis.

With dynamical fermions included no value for the surface tension
has so far been extracted. However,
the lattice calculation indicates that the phase transition
is weaker than in the quenched approximation.
A strong increase in the surface tension seems not to be likely, but the
latent heat can drop significantly with the consideration of dynamical quarks.
However, it has not been demonstrated yet that this,
together with the decrease of the transition temperature, could lead to an
increase in the scale of inhomogeneities, which would then affect primordial
nucleosynthesis.
Indeed, it has been argued that this possibility is unlikely \cite{H,BaGa92}.

\begin{center} {\bf Acknowledgement}  \end{center}
This work was supported, in part, by the Natural Science and Engineering
Research Council of Canada.

\end{document}